\documentclass[12pt, preprint]{aastex}
\usepackage{psfig}
\def\gs{\mathrel{\raise0.35ex\hbox{$\scriptstyle >$}\kern-0.6em 
\lower0.40ex\hbox{{$\scriptstyle \sim$}}}}
\def\ls{\mathrel{\raise0.35ex\hbox{$\scriptstyle <$}\kern-0.6em 
\lower0.40ex\hbox{{$\scriptstyle \sim$}}}}

\def\etal{\hbox{et al.}$\,$}
\def\OII{\hbox{[O II]}$\,\,$}
\def\Hd{\hbox{H$\delta$}$\,\,$}

\def\EWOII{\hbox{EW([O II])}$\,$}
\def\EWHd{\hbox{EW(H$\delta$)}$\,$}

\def\Msun{\rm{\hbox{M$_{\odot}$}}}           
\def\ang{\hbox{\AA}$\,$ }
\def\kms{\rm{\hbox{km s$^{-1}$}}}
\def\OII{\hbox{[O II]}$\,$}
\def\HII{\hbox{H~II}$\,$}
\def\Hd{\hbox{H$\delta$}$\,$}
\def\24m{\hbox{24~$\micron$}$\,$}

\newcommand{\microJy}{$\mu$Jy}

\slugcomment{Submitted: 2008 June 13; accepted: 2008 --- ---}

\shorttitle{Dressler et al.}
\righthead{Spitzer-MIPS \24m detections in Abell 851}

\begin{document}

\title{Spitzer \24m detections of starburst galaxies in Abell 851}

\author{Alan Dressler, Jane Rigby, \& Augustus Oemler Jr.}
\affil{The Observatories of the Carnegie Institution of 
Washington, 813 Santa Barbara St., Pasadena, California 91101-1292}
\email{dressler@ociw.edu, jrigby@ociw.edu, \& oemler@ociw.edu}

\author{Jacopo Fritz \& Bianca M.\ Poggianti}
\affil{INAF-Osservatorio Astronomico di Padova}
\affil{Osservatorio Astronomico di Padova, vicolo 
dell'Osservatorio 5, 35122 Padova, Italy}

\email{bianca.poggianti@oapd.inaf.it \& jacopo.fritz@gmail.com0}

\author{George Rieke \& Lei Bai}
\affil{Steward Observatory, University of Arizona, Tucson, AZ 85721}
\email{grieke@as.arizona.edu \& bail@as.arizona.edu,}

\begin{abstract}

Spitzer-MIPS \24m observations and ground-based optical imaging and spectroscopy of 
the rich galaxy cluster Abell 851 at $z = 0.41$ are used to derive and compare star formation 
rates from the mid-IR \24m and from \OII $\lambda\lambda$3727 emission.  Many cluster 
galaxies have SFR(\24m) /SFR(\OII) $>>$ 1, indicative of star formation in regions highly 
obscured by dust.  We focus on the substantial minority of A851 cluster members where 
strong Balmer absorption points to a starburst on a $10^{8-9}$ year timescale.  As is typical, 
two types of galaxies with strong Balmer absorption are found in A851: with optical emission
(starforming), and without (post-starburst).  Our principal result is that the starforming variety, 
so-called e(a) galaxies, are mostly detected (9 out of 12) at \24m --- for these we find typically 
SFR(\24m)/SFR(\OII) $\sim 4$.  Strong Balmer absorption and high values of SFR(\24m)/SFR(\OII) 
indicate moderately active starbursts; both observations support the picture that e(a) galaxies are the 
active starbursts that feed the post-starburst population.  While \24m detections are frequent 
with Balmer-strong objects (even 6 out of 18 of the supposedly ``post-starburst'' galaxies are 
detected), only 2 out of 7 of the continuously starforming `e(c)' galaxies (with weak Balmer 
absorption) are detected --- for them, SFR(\24m)/SFR(\OII) $\sim 1$.  Their optical spectra 
resemble present-epoch spirals that dominate today's universe;  we strengthen this association 
by showing that SFR(\24m)/SFR(\OII) $\sim$ 1 is the norm today.  That is, not just the amount of star 
formation but also its mode has evolved strongly from $z \sim 0.4$ to the present.  We fit 
spectrophotometric models in order to measure the strength and duration of the bursts and to quantify the 
evolutionary sequence from active- to post-starburst.  Our results harden the evidence that
moderately active starbursts are the defining feature of starforming cluster galaxies at $z \sim 0.4$.


\end{abstract}


\section{Introduction}

A remarkable change in our view of galaxy evolution over the last few decades has been the 
realization that the properties of galaxies are changing on relatively short time scales, in cosmic 
terms.  The Butcher-Oemler (1978) effect, first at-odds with the then-prevailing 
picture that strong galaxy evolution occurred only within the first few billion years, is now part 
of the orthodoxy -- a large and still-growing body of evidence shows that the stellar populations
and even the morphologies of many galaxies have undergone significant change, even within the 
last 4 billion years.  Butcher and Oemler's discovery of a substantial population of luminous starforming 
galaxies in intermediate-redshift clusters -- a great contrast with cluster populations today where
only a small fraction of galaxies are still actively forming stars -- is now widely confirmed and accepted.  

Less well known, but likely as important for understanding the causes of the Butcher-Oemler 
effect, is the discovery by Dressler and Gunn (1983) that most of these starforming galaxies in the 
cores of distant clusters show a starburst signature in their optical spectra: both starbursts and post-starbursts 
are common.  Later work has confirmed the ubiquity of this result (Couch \& Sharples 1987, Wirth \etal 1994,
Poggianti \etal 1999 -- P99, Tran \etal 2003).\footnote{Interested readers will appreciate 
a review of the often-cited rebuttal by Balogh \etal (1999).  These authors actually found a 5\% 
incidence of strong Balmer-line sources --- for a lower redshift cluster sample, with a higher threshold of 
Balmer strength --- quite consistent with the other work cited above.  Balogh \etal discounted their 
detections as simple noise in their spectra, after determining the ``noise'' from the \Hd distribution in their 
full sample.  Dressler \etal (2004) show that the detections were in fact reliable -- and the noise greatly 
overestimated --- due to a difficulty with the bandpass measurement technique, which for small \Hd equivalent 
width returns negative values --- see also Prochaska \etal 2007.}  Data from these studies suggest
that a significant fraction of field populations at $z \sim 0.5$ is also undergoing moderate bursts 
of star formation  (Dressler \etal 2004).  This behavior contrasts with today's population of field 
galaxies, for which continuous star forming is the dominant mode, and starbursts are rare.  This is yet 
further evidence of rapid evolution of the galaxy population, at least as important, since it applies to field 
and cluster galaxies alike. 

In this paper we address one of the long-standing uncertainties of this subject --- the amount of 
dust-obscured star formation that has not been generally accounted-for in representative field
and cluster populations at $z \sim 0.5$. Although Balmer absorption is a reliable indicator of past
starburst activity, a coherent picture requires evidence of the bursts themselves. This has been
hampered by the possibility that much of the star formation is obscured by dust, and 
therefore not part of the tally made with optical-to-near-IR observations. The remarkable 
sensitivity of the Spitzer Space Telescope with its \24m MIPS camera is allowing great progress
in removing this uncertainty. 

As part of a successful Spitzer program to observe two rich clusters of galaxies, we investigated
archival MIPS data for Abell 851, a cluster that is well studied in the optical, x-ray, and radio, and for 
which a significant spectroscopic sample has been obtained and explored in a companion paper 
(Oemler et al. 2008, hereinafter Oem08).  Our purpose here is limited to the implications for the amount 
and character of star formation in distant clusters, specifically the subject  of ``starburst vs. continuously 
starforming'' for distant galaxies.  Issues of causes and mechanisms are beyond the scope of this paper, 
but Spitzer data for additional rich, distant clusters and their surrounding field populations will be able to 
advance these discussions considerably.

The paper is organized as follows.  In section 2 we lay out the issues, in Section 3 we address the 
ability of mid-infrared imaging to add to our knowledge of the star formation rates of distant galaxies, 
in Section 4 we present the optical and mid-IR data for A851, and Section 5 we give the results of 
combining these data sets.  In Section 6 we compare our detections of dusty starbursts in A851 with
similar observations of present-epoch spirals.  Finally, in Section 7, we apply stellar population
models to the optical spectra + MIPS data and discuss the results in terms of a consistent picture
of the starburst activity in distant clusters.

\section{Where are the starbursts that feed the post-starburst population?}

Dressler \& Gunn (1983)  first reported spectra of L* galaxies in distant clusters with 
strong Balmer absorption but no emission lines and interpreted these as {\it post-starburst} 
(see also Couch \& Sharpes 1987, P99).  The question ``where are the {\em active} starbursts that
turn into post-starburst galaxies?" was largely set aside.  Dressler, Gunn, and
Schneider (1985) noted only a few examples in CL0024+24 of nearly pure emission-line
spectra --- characteristic of today's strong starbursts, for example, Markarian galaxies.  
But, because it was expected that the starburst phase would be short, $\tau \ls 10^8$ yr, 
compared to the $\tau \sim 10^9$ yr that the galaxy would be recognized as post-starburst, 
it was reasonable at first to expect to catch few of the actual starbursts. This explanation 
became untenable as the Morphs study of 10 clusters accumulated more than 100 post-starburst
spectra: although some tens of emission-line-dominated spectra were identified,
their distribution was distinctly shifted to lower-luminosities than the post-starburst
galaxies --- the predecessors to the post starbursts would be, if anything, more luminous (P99).

However, the larger Morphs sample provided an alternate explanation.  In addition to the post-starbursts, 
the Morphs sample included a substantial number of galaxies with strong \Hd\ {\it\bf and} (relatively weak) 
\OII emission.  Initially, these were interpreted as decaying starbursts --- a phase of subsiding star 
formation between the main burst and the post-starburst phase.   However, the dusty appearance of 
many of these `e(a)' galaxies (`e' for emission, `a' for A-stars) led P99 to another explanation --- that 
e(a) galaxies were, in fact, the 
actual starbursts.  If  heavily obscured by dust, \OII emission would greatly under-represent the star 
formation rate.  As examples, P99 pointed to the Liu \& Kennicutt (1995) sample of present-day mergers, 
which shared the spectral properties of the e(a) galaxies (see Poggianti \& Wu 2000, Poggianti,
Bressan, \& Franceschini 2001).  Liu \& Kennicutt showed, from IRAS mid-IR observations and 
optical imaging, that the majority of star formation in such systems is hidden by dust
from optical or near-IR observation.  Detection with the VLA of a radio continuum of many of these 
putative dusty starbursts in the distant cluster sample also supported the idea of hidden star 
formation (Smail \etal 1999).  Furthermore, radio detections of a few supposedly {\it post-starburst} 
galaxies suggested that dust might be hiding continuing star formation in some of these as well.

Subsequent work has been supportive of the identification of dusty starbursts in distant clusters.
ISOCAM contributed extensively to the study of mid-IR emission from present-epoch galaxies at all types 
was extensive.  In particular, Duc \etal (2002) and Fadda \etal (2002) made pioneering observations 
of galaxies in clusters at $z \sim 0.2$.  ISOCAM's sensitivity did not allow it to detect typical starforming 
galaxies at $z \sim 0.5$, and even at $z \sim 0.2$ the detections were mostly luminous infrared galaxies (LIRGs).  
Despite limited sensitivity, several studies detected luminous, dusty starbursts in clusters at these 
redshifts, galaxies that are rare (at best) in rich clusters at z=0.  The pre-Spitzer era of mid-IR emission from 
galaxies is well and thoroughly reviewed by Metcalfe \etal (2005).  

Now, with MIPS on Spitzer, it is possible to measure star formation rates $\sim1~\Msun~yr^{-1}$
at $z \sim 0.5$, a value typical of normal spiral galaxies today.  This means that L* galaxies with even 
modestly elevated rates of star formation compared to today's spirals are detectable, allowing us to 
test directly the claim by the Morphs that the e(a) galaxies in intermediate-redshift clusters are dusty 
starbursts of sufficient luminosity to feed the post-starburst population and judge, as well, whether the 
post-starburst galaxies are really {\bf post} starburst.

Ours is not the first study to use MIPS on Spitzer to image intermediate-redshift clusters.  Geach \etal 
(2006) targeted two clusters at $z \sim 0.5$ and found many mid-IR sources in one, although 
comparatively few in another cluster of similar richness.  They describe the properties of the 
population -- luminosity function, spatial distribution, etc., but do not analyze the properties the 
individual galaxies and basically bypass the question of starbursts, consistent with the paradigm 
favored by Geach \etal in which starforming galaxies falling into clusters are simply extinguished.

Bai \etal (2007) and Marcillac \etal (2007) imaged two very rich clusters at $z \sim 0.8$, for which they
detected 30-plus galaxies in the mid-IR that are either confirmed or suspected (photometric
redshift) cluster members.  Because of the greater distance relative to the Geach \etal clusters
and A851, the detected galaxies have greater rates of star formation --- tens of solar masses 
per year --- most of these can be classified as LIRGS.  The optical spectra of these mid-IR 
detections argue that these galaxies are strongly star-forming and considerably obscured by dust.
Both Bai \etal and Marcillac \etal,  like Geach et al., analyze the properties of the population, velocity 
kinematics,  galaxy morphology, spatial distribution, particularly with respect to x-ray emission, colors, 
and luminosity distribution.   In addition, Marcillac \etal investigate the correlation with optical spectral 
types -- passive, starforming, and post-starburst, but their sample included only 4 galaxies that are 
candidates for active, dusty starbursts, of which 3 are detected in the mid-IR.  Our goal in this and 
subsequent studies is to extend this comparison of optical spectra and mid-IR emission to typical 
starforming galaxies of lower luminosities, which will provide a fuller picture of the evolution of the 
cluster population, $L > 0.4L*$.

This study of A851 and the future observations we will obtain with Spitzer-MIPS will provide crucial 
data for understanding the star formation history of galaxies that are involved in building 
intermediate-redshift clusters.  The optical spectroscopy and photometry of such samples are missing 
information about dust obscuration that is necessary to this task; adding the Spitzer data provides 
such data.  We will pay particular attention to the presence of Balmer absorption in 
our sample galaxies, highlighting the importance of starbursts in cluster evolution 
(see D99, P99 and references therein).  Unlike the other studies of clusters with Spitzer-MIPS, 
we will concentrate on modeling the star formation history with the critical addition of the mid-IR data.  
Specifically, through basic stellar population models of the data, we will address the question of 
whether the distinctions we have drawn between continuously starforming, starbursting, and 
post-starburst galaxies are quantitatively consistent with the observations of intermediate-redshift 
cluster galaxies.

\section{The Data}

\subsection{Basic properties of the rich cluster of galaxies Abell 851}

The rich cluster Abell 851 (aka CL0939+4713) at $z = 0.41$ is one of the best-studied 
intermediate-redshift clusters and one of the clusters in which post-starburst galaxies 
were first identified (Dressler \& Gunn 1992).  A851 is among the most populous clusters 
known at its epoch and is also noteworthy because it appears to be in the process 
of assembly through the mergers of several subclusters. This is especially evident in X-ray 
observations (Schindler \etal 1998) where two prominent centers of emission are also 
centers of galaxy concentration.

Although A851 is unusually rich and dynamically active, its galaxy population is not
qualitatively different from that of other rich clusters at $z \sim 0.5$, as shown by studies
by the Morphs collaboration (Smail \etal 1997, Dressler \etal 1999, P99, D04, Oem08).  The 
majority of its galaxies are red, passive elliptical and S0 galaxies, but A851 is also well 
represented in both types of galaxies with strong Balmer absorption: those without emission 
that are identified as post-starburst,
and those with emission that have been suggested to be dust-obscured, 
active starbursts that feed the post-starburst population (P99).  

Oem08 studied an extended region $R \sim 2-3$ Mpc around the core of A851, including HST 
images and spectroscopy for both an extended square region and the remarkable filament that 
stretches further, to the north-west of the cluster.  Oem08 found a substantial population of  
active starbursts and fewer post-starburst galaxies in the outskirts of A851, consistent with
idea that many of the starbursts are triggered well outside the cluster core, probably within 
the infalling groups from which the cluster is growing (P99, Treu \etal 2003, Moran \etal 2007).  
Oem08 also found another population of starbursts that appear to have been triggered by 
passing through the cluster core. 

Oem08 also show that the morphology-density relation for this cluster is well developed, even 
though it is, at $z = 0.41$, in a unrelaxed dynamical state.  From the spatially extended sample it 
appears that A851 is one of the most active clusters known in terms of star formation and starbursts, 
but that these differences are quantitative rather than qualitative, and likely the result of the very 
dynamic phase in which it is observed.  A complete description of A851, the observations, and the 
implications for galaxy evolution in clusters can be found in Oem08.

\subsection{Optical data and Spitzer-MIPS Observations}

The optical data for this study includes photometry and spectroscopy from Dressler \& Gunn (1992)
and the Morphs collaboration.  HST WFPC-2 observations that cover a 3x3 mosaic (480 arcsec square)
are described in Smail \etal (1997) and Oem08.  The morphological information from these images
is mainly discussed in Oem08 and briefly referenced here, but these fields basically define the 
extent of the photometric and spectroscopic sample that we use here. The photometric data
come from these HST frames and ground-based observations by Dressler \& Gunn. The spectroscopic 
data come from Dressler \& Gunn and the Morphs collaboration (Dressler \etal 1997, D99); they are 
explained in more detail in Oem08. 

The core of Abell 851 was mapped in 24~\micron\ by MIPS in Spitzer GTO program 83 
(PI G.~Rieke).  The background level was 32 MJy/SR, which is classified as ``medium'' by 
the SSC.  The observation was done in Photometry Mode, with six cycles of 
half-frame--overlapping coverage, and individual exposure times of 30~s.  The resulting
exposure time per-pixel is 2685~s in a 5\arcmin\ $\times$ 5\arcmin\ box centered on the 
cluster, and 900~s in two flanking boxes each 5\arcmin\ $\times$ 2.2\arcmin\ in area.  The 
online SSC tool SENS-PET estimates that these exposure times should have $5 \sigma$ 
point source sensitivities of 70 and 120 \microJy, respectively.  At these depths, 
confusion noise from extragalactic sources becomes important.

The 24~\micron\ images were reduced and combined into a mosaic using the 
Data Analysis Tool (Gordon \etal\ 2006) with a few additional processing steps 
(Egami \etal\ 2006).  Photometry at 24~\micron\ was obtained by PSF fitting
using the DAOPHOT task \textit{allstar}, as detailed in Rigby \etal (2008).

The areas covered by optical spectroscopy and MIPS 24 micron are of similar size, but not 
entirely coincident.  There are 101 cluster members with high--quality optical spectra ($Q \le 3$, see 
Oem08 for more information).  Of these, 83 have MIPS \24m coverage.  Of those, 22 are 
Spitzer detections with $f(\24m) > 80~\mu$Jy.   Using a $R=1.44$\arcsec\ matching 
radius, and given the number density of \24m sources with $f > 80~\mu$Jy in the image, we expect 
from the P-statistic (Lilly et al. 1999) that approximately one cluster member will be spuriously matched 
to a \24m detection with $f>80~\mu$Jy .  For non-detections in the regions of deep and shallower 
coverage we quote upper limits of $f < 80~\mu$Jy and $< 120~\mu$Jy, respectively.  For non-detections
located close to bright sources, we assign a larger upper limit of $f < 200~\mu$Jy.

\section{Measurements of star formation histories from optical and mid-IR data}

\subsection{Optical diagnostics}

In this study we focus on the emission feature \OII $\lambda\lambda$3727 oxygen doublet and the
Balmer absorption line \Hd as indicators of the star formation rate (SFR).   As discussed in P99 and 
D04, these two features measure the time-averaged SFR over $\tau \sim 10^7$ yr and $\tau \sim 10^9$ yr, 
respectively.  Measurements of \OII and \Hd from our sample spectra have been done with a 
semi-automated line-fitting technique, described in Oem08.  A quantitative comparison of the 
strength of these indicators measures whether the star formation rate has been steady over the 
previous $\tau \sim 10^9$ yr or whether there has been a starburst with a rapid decline in the SFR 
rate over that timescale.  An ongoing starburst is inferred by comparing the current ($\tau \sim 10^7$ yr) 
SFR to the total stellar mass of the galaxy, through spectral synthesis modeling.

In the following we use the strengths of \OII and \Hd to divide galaxies into spectral types which correspond 
to different histories of star formation (see Oem08, Table 5 for the definition of the types); e(c)'s are galaxies 
with continuing star formation, k+a's and a+k's are moderate and strong post--starbursts, e(b)'s are starbursts with 
strong optical emission, e(a)'s are dusty starbursts with weak optical emission, k's are passive, early--type 
galaxies, and e(n)'s are AGN's. By starbursts, we shall mean any galaxy whose observed star formation rate is 
at least a factor of two higher than its long-term average. We shall use the term ``post--starburst'' to signify any 
galaxy which had a starburst in the recent ($\tau \ls 10^9$yrs) past, but at the epoch of observation shows no 
detectable star formation, and buried starburst to signify a starburst which is sufficiently obscured by dust to 
completely hide its \OII emission.

Oem08 have measured the equivalent width of \OII in the A851 sample. We convert these equivalent widths 
to \OII luminosities by the following steps. (1) Because our optical spectra cover only a fraction of each galaxy, 
while the Spitzer photometry covers the entire object, we must make some assumption about the distribution of 
\OII flux over the face of the galaxy. In the absence of any other information, we assume that the \OII flux has the 
same distribution as the visble light. (2) Using the total r magnitudes of the galaxies, we calculate the total 
$F_\nu$ at 6500\AA. (3) Using the shape of our fluxed spectra, we then determine the total $F_\nu$ at 3727\AA, 
and from this and the measured \EWOII, we calculate the total flux in the \OII line. (4) We then use a standard 
$\Lambda$ cosmology to calculate total \OII luminosity. This method is dependant on our assumption about the 
spatial distribution of \OII, but the spectra, which come from slits that typically cover $\sim8 \times 12$ kpc at 
$z = 0.41$, sample a reasonable fraction of each galaxy in our sample, so the sensitivity of our result to this 
assumption is probably not severe.

The UV flux from young, massive stars in \HII regions ionizes oxygen atoms which then produce \OII 
through recombination. Only the hottest stars have sufficient UV flux to contribute, so \OII flux in an 
\HII region is due mainly to the population of O and B stars.  Rates of star formation come from
folding in the well-known lifetimes of these stars.  Deriving the SFR
for stars of all masses requires the adoption of a universal initial mass function --- an 
often-questioned, but yet to be invalidated, assumption.  We have employed Kennicutt's (1998) 
prescription:

{\noindent SFR $\Msun yr^{-1}$ = $1.4 \times 10^{-41}$ L(\OII) ergs~s$^{-1}$      (1)}

{\noindent We note that, because this is an average rate that includes lower-luminosity spiral and irregular 
galaxies less luminous irregular galaxies, we are perhaps overestimating the SFR for our 
sample --- by $\ls40\%$ --- which makes this a conservative choice for the purposes of this study.  
On the other hand, there might be even greater uncertainty just from using this $z \sim 0$ relation 
on earlier-epoch starforming galaxies.  We show below that, despite these uncertainties, we derive 
approximately equal SFRs from both the Kennicutt \OII relation and from \24m for ``normal" 
starforming galaxies (in both A851 and a present-epoch field sample covering the same luminosity 
range), which is reassuring.} 

The Kennicutt formulation is sometimes corrected for dust extinction and sometimes not.  
The relation adopted here is corrected for average dust extinction from \OII to H$\alpha$, 
but not for extinction at H$\alpha$.  Our goal is to compare the SFRs derived from \OII
to the SFR derived from the dust-insensitive \24m luminosity, so measuring \OII 
flux and applying the standard correction to H$\alpha$ for present-epoch galaxies will show 
whether the distant starforming galaxies in A851 have an unusual amount of dust and thus 
more star formation than is evident with optical diagnostics.  Because our analysis depends
on the relative values of these two indicators, our conclusions are not dependent on a
rigorous analysis of the absolute value of extinction in our sample galaxies, which is beyond
the scope of this paper.

Measurements of \Hd strength are comparatively insensitive to dust aborption, because, over 
the $10^9$ yr timescale for which it measures the SFR, the A-F stars that contribute most 
to the \Hd\ signal are expected to have diffused out of the dusty regions in which they were born.

\subsection{Mid-infrared diagnostic}

Empirically, the total infrared (8--1000~\micron) luminosity, L(TIR), correlates reasonably well with 
other star formation rate tracers in nearby dusty star--forming galaxies (Hunter \etal\ 1986, Lonsdale, 
Persson, \& Helou 1987, Kennicutt 1986). Far-IR emission arises from grey-body emission of 
dust heated by hot stars.  Since the escape fraction for ionizing photons is low, and the
dust cross-section peaks in the UV, the far-IR is essentially a calorimeter of hot young stars 
(Kennicutt 1986, and references therein).  Only at low redshift is it currently possibly to
sample fully the long--wavelength SED of galaxies to measure L(TIR) directly.  Fortunately, 
monochromatic luminosities in the rest-frame mid-IR (5--30~\micron) range correlate well with 
L(TIR) at low redshift (Chary \& Elbaz 2001, Calzetti 2007), and this correlation has been 
demonstrated to extend to $z=1.3$ (Marcillac \etal\ 2006).

Mid-IR emission is a combination of continuum from small grains plus aromatic features. The 
luminosity surface density in both the continuum--dominated $\lambda_r = 24$~\micron\ spectral 
region and the aromatic--dominated $\lambda_r = 8$~\micron\ spectral region correlate well with 
that of extinction--corrected Paschen~$\alpha$ (Calzetti 2007); in both cases the $1 \sigma$ 
scatter is about 0.3 dex.  Figure 3 of Alonso-Herrero et al. (2006) shows a tighter correlation between 
rest-frame 12~\micron\ and extinction--corrected P$\alpha$, $\sigma \sim$ 0.1--0.2 dex.  

At $z=0.4$, the 24~\micron\ band of MIPS on Spitzer detects rest--frame 15--19~\micron.   These 
wavelengths are also dominated by continuum emission from small grains, with a contribution from the 
relatively weak 17~\micron\ aromatic feature. The rest-frame 15--19~\micron\ bandpass has not been 
similarly calibrated at $z = 0$ because IRAS had no such band, however, since the 15--19~\micron\ 
continuum arises from the same small grains that produce the 12um and the 24um rest-frame continua, 
the scatter in this band should be similar.  We conclude from the results from Calzetti, Marcillac \etal, 
and Alonso-Herrero \etal, and also Smith \etal (2006, 2007) that the continuum emission in this 
region should trace the star formation rate with only moderate dispersion.  Because mid-IR photons are 
not further absorbed by other dust grains in starforming regions, they provide a clear view of the total star 
formation in the system, which \OII and even H$\alpha$ cannot do.  

In the next section we will use template SEDs generated by Chary \& Elbaz (2001) and Dale \& 
Helou (2002) that model the flux distribution from the mid-to-far IR to derive the total SFRs
for our detected galaxies.  The SED shapes of these templates vary smoothly as a function of 
mid-IR luminosity, allowing a reliable correction to bolometric IR emission from observations
over a relatively narrow wavelength range. A small complication is that these templates do
not include $\lambda > 15$~\micron\ aromatic features.  This means that the 17~\micron\ 
aromatic feature, which falls in the \24m band at the $z=0.407$ redshift of A851, will not be 
correctly accounted for.  However, the feature has a typical equivalent  width 
of only  0.36~\micron\  (Brandl \etal\ 2006), and affects the detected MIPS 24~\micron\ flux density
only at the $10\%$ level.  This small systematic effect is not important for our results.

For a sample of 59 SINGS galaxies, Smith \etal\ (2007) observed an intrinsic dispersion 
in aromatic feature strength compared to the dust continuum. Figure 19 of Smith \etal\
quantifies how this spectral variation affects 24~\micron\ flux density: at $z=0.4$ the 
$10\%$--$90\%$ variation in the spectra of their luminous (L(TIR) $> 2.6\times 
10^{10}$~L$_\odot$) sample causes a $\pm 20\%$ change in the MIPS 24~\micron\ flux 
density.  Thus, flux differences of this magnitude should not necessarily be interpreted as 
differences in star formation rate, since they may be due to intrinsic spectral 
variation.

Despite these caveats, the robustness from dust obscuration, the well--behaved nature of the 
mid-IR continuum (Calzetti 2007), and the relative weakness of the 17~\micron\ aromatic 
feature makes it a relatively reliable diagnostic of star formation rate in bursting galaxies.

\subsection{Deriving star formation rates from mid-IR fluxes.}

We converted from observed 24~\micron\ flux density to inferred star formation rate as 
follows:  Chary \& Elbaz have derived template SEDs that model the full emission from
8-1000~\micron\ for galaxies with a wide range of mid-IR luminosity.  We transformed each 
of the Chary \& Elbaz synthetic templates to the cluster redshift and converted L$_\lambda$ 
to f$_\lambda$, assuming a flat cosmology with h$_0 = 0.72$ and $\Omega_\lambda = 0.73$.  
We then computed the 24~\micron\ flux density that would be measured for this template using
 the MIPS response curve.\footnote{http://ssc.Spitzer.caltech.edu/mips/spectral\_response.html}

We integrated each template from rest--frame 8--1000~\micron\  to find the total infrared 
luminosity, and converted that to a star formation rate using equation 4 of Kennicutt (1998).  
The result is a look-up table, at the cluster redshift, of star formation rate as a function of 
24~\micron\ flux density.  For the range of detected MIPS flux densities, the corresponding 
Chary--Elbaz templates were numbers 36 to 75.

This extrapolation from rest-frame 15--19 \micron\ to total 8--1000 \micron\ luminosity necessarily 
depends on the templates assumed.  Since we observe large populations of A stars extending
well beyond the starforming regions in galaxies which, we will show, have had significant, long duration
starbursts, we considered the extra heating that these might provide.  Our modeling included comparing
the contribution to mid-IR luminosity from young populations ($\tau \ls10^7$ yr) to intermediate-age
populations ($\tau \sim10^8$ yr).  From this we concluded that, for a Salpeter IMF and a single-age Starburst 99 
model (Leitherer \etal 1999), the power available in A stars to heat dust is only $\sim$1\% of the power available 
from O and B stars.  Thus, we expect our \24m detections to be dominated by O and B stars, even in 
galaxies with large A star populations.

\section{Results}

\subsection{Detection with Spitzer-MIPS is a strong function of spectral type}

In Figure 1 we show the result of our attempt to measure Spitzer \24m flux densities 
for our A851 cluster sample.  We plot Spitzer \24m flux density versus R-band optical
flux for the 96 confirmed members of A851 (see Oemler \etal\ 2008) that are within the 
Spitzer field. Figure 1 shows that --- even for cluster members, which comprise an essentially 
volume-limited sample --- galaxies are detected over a wide range of luminosity.  In 
other words, we are not picking off only the most luminous objects.   Most of the cluster
members are not detected; however, this sample includes many galaxies that are not 
starforming (passive), for which detection by Spitzer at \24m is not expected, unless the 
galaxy harbors an AGN.

The non-detection of galaxies without star formation is seen more easily in Figure 2, from
the data in Table 1  which break down the detections by optical-spectral classification.
Of the 41 galaxies `k' types that are passively evolving, 
only 2 are detected.  There is nothing in the optical spectra of these galaxies that suggests 
any activity.  It is possible that these could be examples of extremely-dust-hidden star formation, 
as we find below for some of the putative post-starburst galaxies, or AGN, or both.


\begin{figure*}
\vspace*{0.1cm}
\hbox{~}
\centerline{\psfig{file=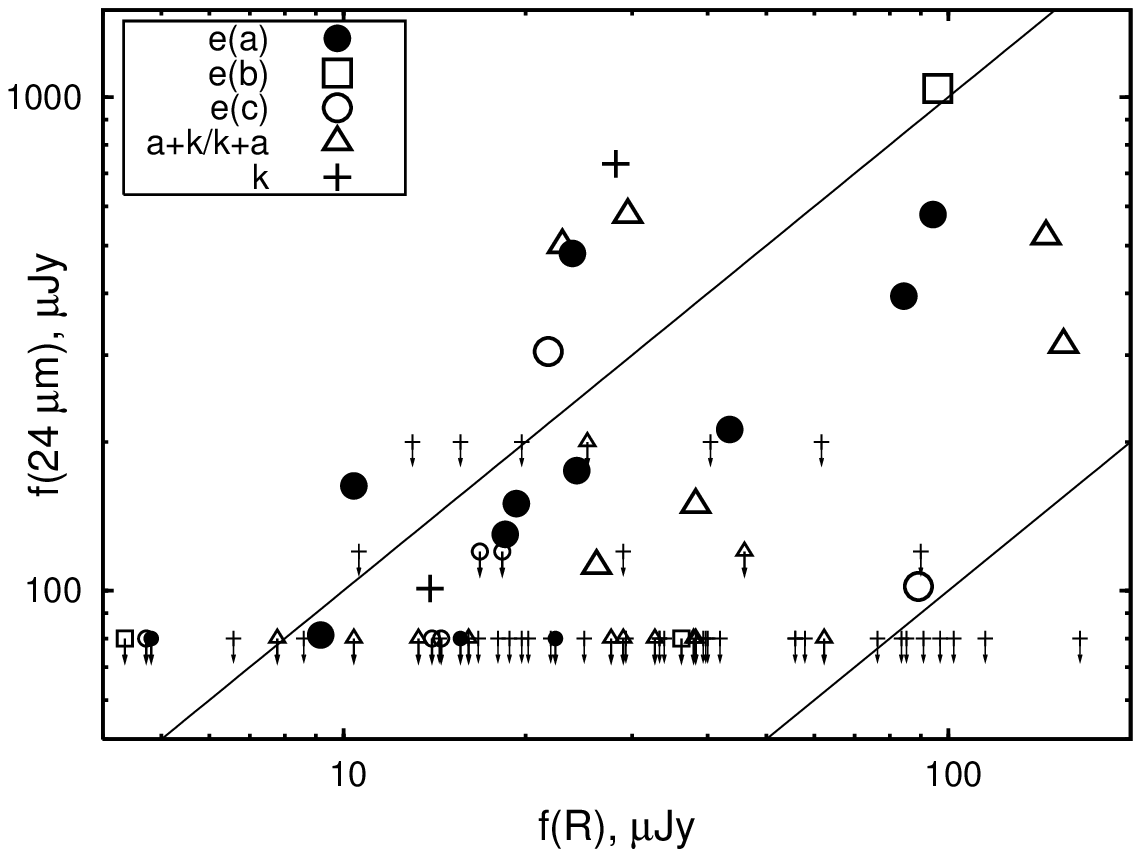,angle=0.,width=6.5in}}
\noindent{\scriptsize
\addtolength{\baselineskip}{-3pt}
 
\hspace*{0.3cm}Fig.~1.\ Spitzer \24m flux density vs, optical R-band flux for the cluster members 
of the optical-spectroscopic sample of A851 galaxies.  Most galaxies are undetected by Spitzer: 
only 2/7 e(c)'s, and 2/41 k's are detected.  (The k-types upper-limits are 
the smaller arrows.)  However, 9/12  e(a)'s, 1/3  e(b)'s, and even 6/18  ``post-starburst'' 
galaxies, k+a \& a+k, are detected, indicating a clear connection between starburst activity and heavily 
dust-obscured star formation.  This figure shows the spread of flux for the sample: a factor of 20 for both 
the \24m observations and the R-band (3.2 mag). 

\vspace*{0.2cm}
\addtolength{\baselineskip}{3pt}
}
\end{figure*}

\subsection{Continuously starforming (``normal'') galaxies are generally not detected}

Our A851 cluster sample includes both continuously starforming (CSF), e(c) galaxies as well as starburst 
(SB) and post-starburst galaxies (PSB).  The strength of Balmer absorption lines in an integrated
spectrum of a galaxy is key to distinguishing the CSF galaxies from the SB and PSB galaxies.
The Balmer absorption line strength is a combination of effects, including:
(1) the dominance of Balmer absorption lines in the spectra of A stars, which contribute most of the 
light in populations of age $\tau \ls10^9$ years; (2) the in-filling of Balmer absorption by Balmer emission 
from the ionized gas in H II regions excited by OB stars; (3)  the migration of A stars from the 
dust-obscured H II regions of their birth, and the distribution and amount of that dust; (4) the 
strength and duration of the starburst; and (5) the relative contribution of light from KIII stars --- 
populations older than $2 \times 10^9$ years.  It is the combination of these effects that results in 
substantially different spectra for CSF as compared to SB or PSB galaxies, the key to distinguishing 
CSF galaxies from other types. 

The part of the A851 cluster sample that is coincident with Spitzer MIPS observations contains 
7 CSF e(c) galaxies.   Only 2 of these are detected with MIPS. The 7 e(c) galaxies cover the full range 
of optical and \24m luminosity, so this  poor detection rate is not mainly due to some luminosity-dependent 
selection effect.  As we show in Section 6, we expect  SFR(IR) $\sim$ SFR(\OII) for these galaxies (from 
a comparable low-redshift,  local sample), which would give them \24m luminosities scattered around the 
Spitzer--MIPS detection limit.


\begin{figure*}
\vspace*{0.1cm}
\hbox{~}
\centerline{\psfig{file=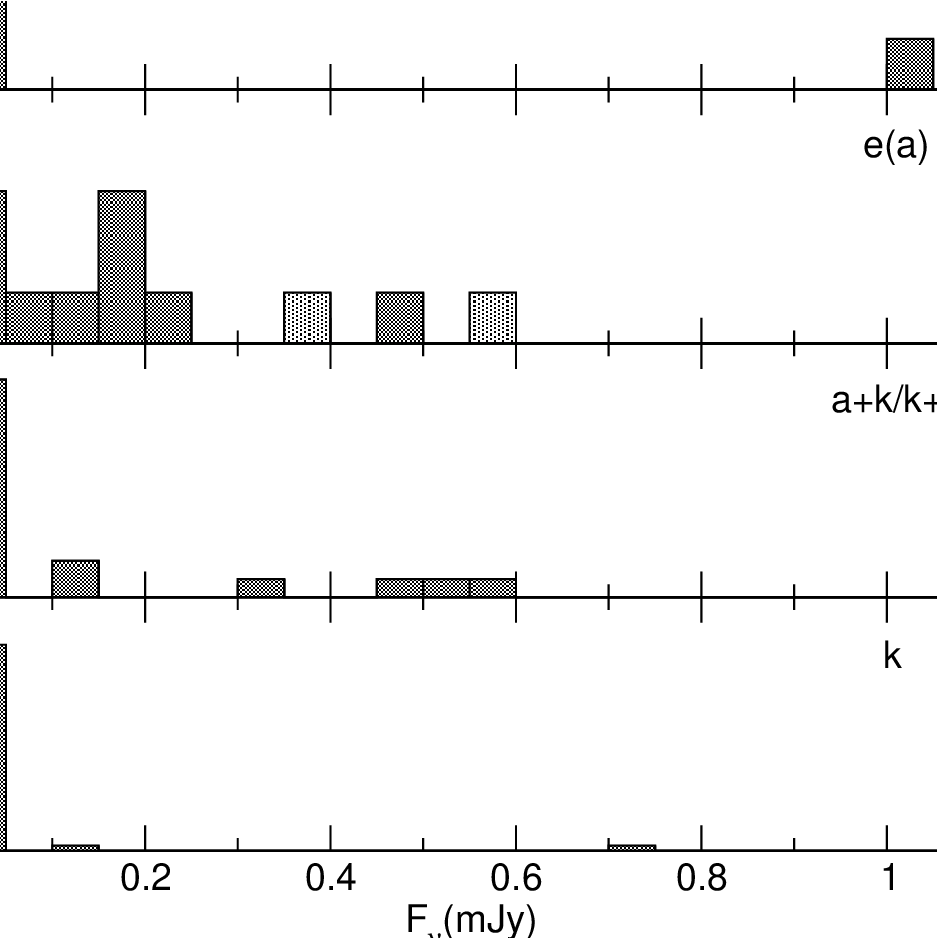,angle=0.,width=4.0in}}
\vspace*{1.0cm}
\noindent{\scriptsize
\addtolength{\baselineskip}{-3pt}  

\hspace*{0.3cm}Fig.~2.\ The distributions of` ``detectability'' as a function of the spectral type.
The horizontal axis is labeled in \24m flux density (below) and star formation rate (above).
The only type for which a majority of galaxies (9/12) is detected is the e(a) class --- the putative dusty
starbursts. (The two lightly shaded squares are k+a galaxies with \OII $\approx~ -4$ \ang, slightly below the
threshhold for the class defined in P99.) The least detected are k-types (2/41) --- passively evolving 
early-type galaxies.  Most post-starburst galaxies are not detected, as expected, but 6/18 are, suggesting that 
some members of this class are still actively forming stars well hidden by dust.   Only 2/7 of the e(c) 
galaxies; some may lie just below the detection limit (but this could also be the case for the 3 undetected
e(a) types).  Of the 3 e(b) galaxies in this sample, only 1 is detected.  Forgetting the small-number statistics, 
this might seem surprising, however, the one detected e(b) is a high-luminosity system, while the others 
are much less luminous, and likely less dusty (as evidenced by the strong emission lines).

\vspace*{0.2cm}
\addtolength{\baselineskip}{3pt}
}
\end{figure*}

\subsection{One-third of post-starburst galaxies are detected}

The post-starburst galaxies of types ``k+a" and ``a+k"  are also mostly undetected, but the 
fraction 6/18 that are detected is significantly higher than even that of the e(c) galaxies, so this 
is a surprisingly high rate for galaxies that are supposed to be observed {\it after} star formation has 
completed.  The fact that some of these post-starburst galaxies could have substantial star formation hidden 
by dust, particularly in their nuclei, had been previously suggested by detections in the radio continuum 
(Smail \etal 1999), preferentially of objects with the strongest \Hd.  These a+k galaxies also show 
the broadest \Hd absorption lines, indicative of the youngest A-star  populations (closest in time to the 
burst) --- see D04 and Oem08.  Of the a+k galaxies in the Spitzer field, 2/5 are detected, compared to 
only 4/13 of the k+a, but with such a small sample, this is only suggestive.

The Spitzer data, then, lead to two important results for the post-starbursts: (1) Most of the 
galaxies classified as post-starburst are exactly that --- there is little or no ongoing, hidden 
star formation (SFR $<$ 3 \Msun\ $yr^{-1}$).  This removes a critical uncertainty in earlier work. 
(2) A significant minority of cluster galaxies classified as post-starburst hide some residual star 
formation. A few of these may be as active as the starbursts we discuss next, but as we show in 
Section 7 (where we model the history of star formation), for most the rate of residual star 
formation is much lower than the preceding burst.  It is quite possible that this 
remaining dust-obscured star formation is concentrated to the center of the galaxy --- this 
is perhaps the only location where a very high extinction can be supported --- while the 
previous, larger starburst  might have been  a more global, and thus more powerful event.  


\begin{figure*}
\vspace*{0.1cm}
\hbox{~}
\centerline{\psfig{file=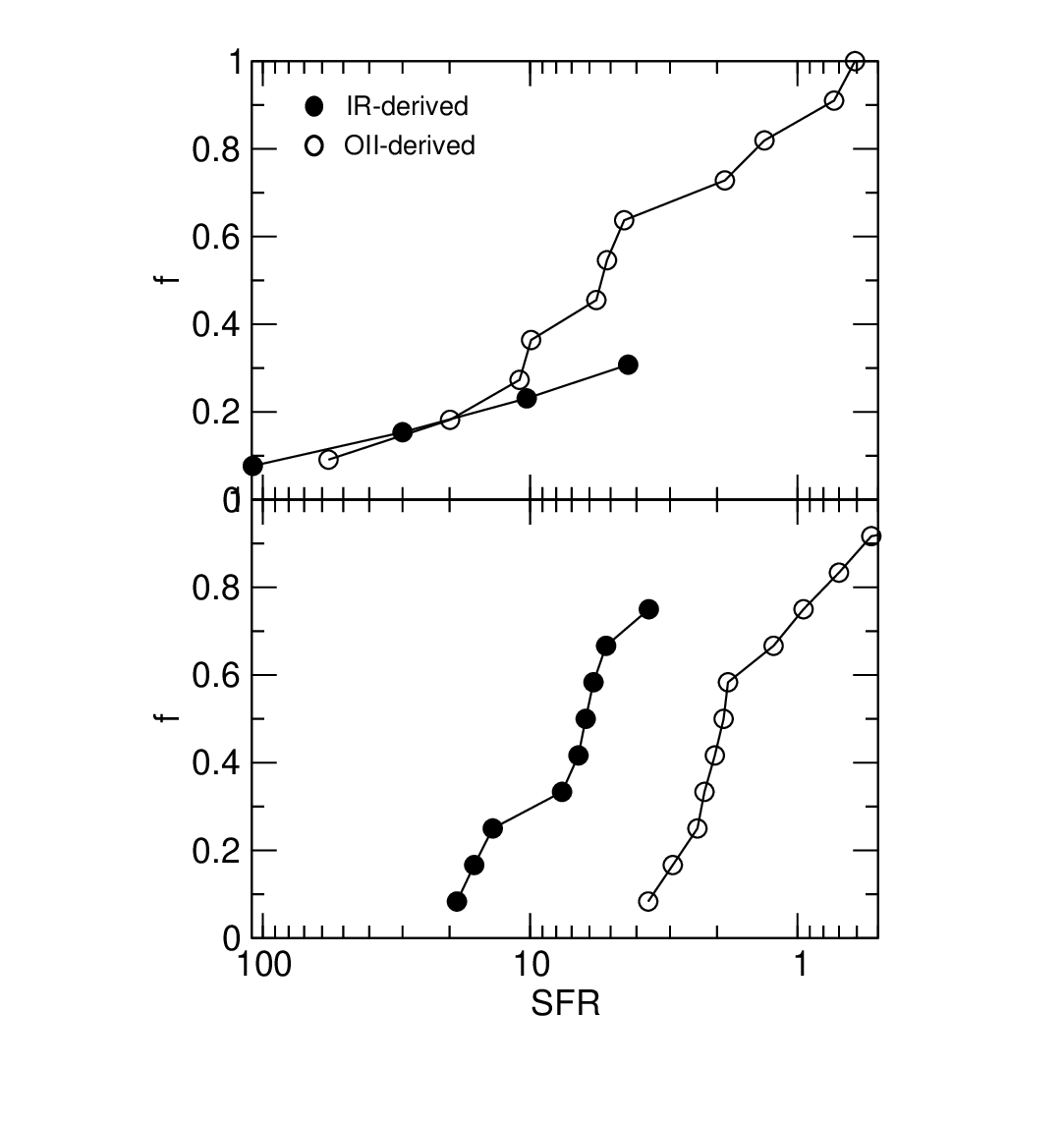,angle=0.,width=6.5in}}
\noindent{\scriptsize
\addtolength{\baselineskip}{-3pt}  
\hspace*{0.3cm}Fig.~3.\ The cumulative distributions of star formation rates (SFRs) for the galaxies in A851 
measured from Spitzer \24m fluxes and \OII emission lines.  The data are for all galaxies with 
EQW(OII) $< -4$ angstroms. Galaxies in which EQW(H$\delta) < +4$ \ang --- e(c) ---  are shown in the upper 
plot; galaxies with H$\delta > +4$ \ang --- e(a) --- are shown on the bottom.  The weak Balmer absorption of the 
upper sample is indicative of normal, continuously starforming galaxies, while the bottom is a sign of starburst 
activity.  The line with open circles shows the cumulative distributions of SFRs derived from \OII, and the 
line with solid points shows the distribution with SFRs dervied from \24m.  The lines trace each other in 
the upper plot --- both methods measure the same SFR, but in the starbursts of the lower plot show an 
offset that corresponds to a factor of 4 greater SFR(\24m) compared to the SFR(\OII).  In other words, 
starburst galaxies --- as identified by \Hd absorption --- have a majority of their star formation obscured by 
dust, while continuously starforming galaxies do not.

\vspace*{0.2cm}
\addtolength{\baselineskip}{3pt}
}
\end{figure*}

\subsection{Candidate ``active starburst" galaxies -- the majority are detected!}
As explained in Section 2, our hypothesis has been that e(a) galaxies, which show both strong \Hd and 
some \OII, are active starbursts that feed the post-starburst population.  Because the \OII\ strength of
these systems is insufficient to indicate a starburst, this only makes sense if most of their star formation 
is hidden by dust.  The Spitzer \24m data for A851 confirm that hypothesis --- for this cluster, at least.  
For the e(a) galaxies, 9/12 are detected.\footnote{This includes 2 galaxies that we detected in \OII with 
an equivalent width of \OII $\sim4$ \ang. This would have placed them in the k+a category previously, 
but in the context of studying SFR from \OII, it makes more sense to include them as e(a) galaxies here, 
because their weak [O II] flux is the result of extinction, not a low star formation rate.} This is our major 
result.

To quantify what this means for the amount of star formation that has been dust-obscured, we 
calculate independently SFRs from \OII and from \24m, as described in Sections 4.1 and 4.2. Because 
there is considerable uncertainty in the measurements for each individual galaxy (the calibrations will depend
on average values of dust temperature and UV-excitation by young stars, for example), we show in Figure
3 the result of this comparison as a cumulative distribution of SFRs for both indicators.  We plot
only the galaxies for which \OII is detected, $\OII~ \le -4$ \ang.  The upper panel is for weak \Hd absorption, what
we identify as CSF --- spectral type e(c) --- starting at 1 \Msun~$yr^{-1}$ in the upper right corner and proceeding 
to very high rates of star formation for a few galaxies in the 10's of \Msun~$yr^{-1}$.  The solid points show 
the rates inferred from the \OII indicator, the open circles, the \24m indicator.  The fact that the two lines
closely follow each other means that approximately the same star formation rate is measured with both indicators:
SFR(\24m)/SFR(\OII) $\sim 1$.\footnote{Because of the inclusion of a partial extinction correction in the 
Kennicutt relation for SFR(\OII), SFR(\24m)/SFR(\OII) $\sim$ 1 occurs at an extinction of about 0.9 mag
at \OII, calculated for a foreground screen (see Kennicutt 1998).  Equality of these two derived star formation rates 
does not mean that star formation is completely unobscured, but is attenuated by dust as is typical for a 
present-epoch spiral galaxy.}

The situation for the \Hd-strong, starbursting galaxies is very different.  The two lines track each other, 
but with a clear offset: SFR(\24m)/SFR(\OII) $\sim 4$. In other words, the \24m flux indicates an SFR that is, on-average,
a factor of 4 greater than is indicated by the \OII emission line, a sufficiently greater SFR that these qualify
as the active, moderate starbursts that feed the post-starburst galaxy population. In the next section we show that 
SFR(\24m)/SFR(\OII) $\sim 1$ is typical of local, present-epoch, continuously starforming galaxies, 
whereas those with SFR(\24m)/SFR(\OII) $\sim 4$ --- the low end of today's LIRGs --- are relatively rare.

\section{Comparison with local samples of starforming galaxies}

Based on the \OII and \Hd spectral indices, P99 and D04 have argued that moderate starbursts 
--- like those in A851 we are discussing here --- are a common feature of the intermediate-redshift 
galaxy population, both cluster and field.  The Spitzer--MIPS data for A851 give us an opportunity
to investigate this claim and remove the uncertainty of dust-obscured star formation to which the
optical indicators are more prone. In order to demonstrate that the prevalence of dust-obscured starbursts 
at $z \sim 0.4$ is an epoch-dependent effect, we need to compare to a like sample of starforming 
galaxies at the present epoch.  Do present-epoch starforming galaxies mostly have SFR(\OII) $\sim$
SFR(IR), indicating that the modest extinction assumed in deriving SFR(\OII) applies? Or, do many
of them have a majority of their star formation hidden by dust, so that SFR(\24m) $>>$ SFR(\OII)?  

Despite the proliferation of local galaxy surveys, constructing a local comparison sample is not easy.  
To begin with, we cannot use present-epoch rich clusters as a source, since these contain very few 
starforming galaxies of mass $M > 10^{10}~\Msun$; this is the Butcher-Oemler effect, the point at 
which we started. We must instead use galaxies in the field --- isolated and in groups --- to judge 
if  the characteristic {\it mode} of star formation, as well as its amount, has changed over time.

Another difficulty is that we need both \24m observations and \OII~ fluxes that cover the full disk of each
galaxy in order to correspond to the distant cluster data.  The survey that comes closest to this is a study 
of 417 starforming galaxies by Moustakas \& Kennicutt (2006).  In this study, \OII~ fluxes were obtained 
from observations for which the slit of the spectrograph was scanned across the galaxy in order to obtain 
an integrated \OII flux, as we obtain by necessity for intermediate-redshift galaxies by using $\sim$1 
arcsec-wide slits.  Unfortunately, this survey is not volume limited, but rather it was chosen to
represent many different galaxy types: not only normal galaxies, but also starbursts, peculiar galaxies, 
interacting/merging systems, and dusty, infrared luminous galaxies were handpicked for inclusion. 
As a result, we were not able to use this study to assemble a sample, based on optical luminosity alone, 
that was complete in any volume-limited sense.

However, the Kennicutt-Moustakas survey provides data that does allow us to construct such a 
sample.  In Figure 4 we show total U-band flux  vs integrated \OII~ flux for all galaxies from that survey 
for which both measurements are available.  Fluxes are from the tables of Moustakas \& Kennicutt (2006), 
which for U-band is generally taken from RC3 (deVaucouleurs et al. 1995), and correspond to
total magnitudes, that is, photometry through fixed apertures has been used to extrapolate a total
brightness for the galaxy in the U-band.  The strong correlation is due to the fact that the UV flux arises from 
the hot young stars that power \OII emission from the H II regions.  Even with the ambiguity of dust 
absorption/obscuration, the fact that these two indicators arise from the same spatial regions assures 
a good correlation.  From the data in Figure 4, we fit a relation between integrated U-band photometry and 
integrated \OII\ emission, so that we can use U as a proxy for OII:

log L([O II]) = M(U-band)/-2.5 +32.8404  (2)

Next, we construct a volume--limited local sample of galaxies using the NASA/IPAC Extragalactic Database 
(NED), as follows.  We use an ``Advanced All-Sky'' NED query, restricting the velocity to  
$1000~\kms~ < v < 2500~\kms~$, and select the 241 galaxies with absolute magnitude $M_V < -19.5$.  We 
remove galaxies whose morphological type is listed as E or S0.\footnote{This large percentage reflects the 
prominence of the Virgo Cluster.} For the remaining 133 galaxies, we took photometric measurements from NED:  
112 had a total U-band measurement (all but 5 from the RC3), and 94 were detected by IRAS at 25~\micron.  
(The several-arcminute beamsize of IRAS means that the mid-IR flux is also an integrated quantity.) We convert 
U and 25\micron\ fluxes to absolute U-band magnitude and monochromatic 25\micron\  luminosity using the 
velocities listed in NED, assuming pure Hubble flow and $H_o = 72$.  Since the IRAS survey was all-sky, and
85\% of the sample has tabulated U-band data, this sample is for the present purpose volume limited and unbiased.

\begin{figure*}
\vspace*{0.1cm}
\hbox{~}
\centerline{\psfig{file=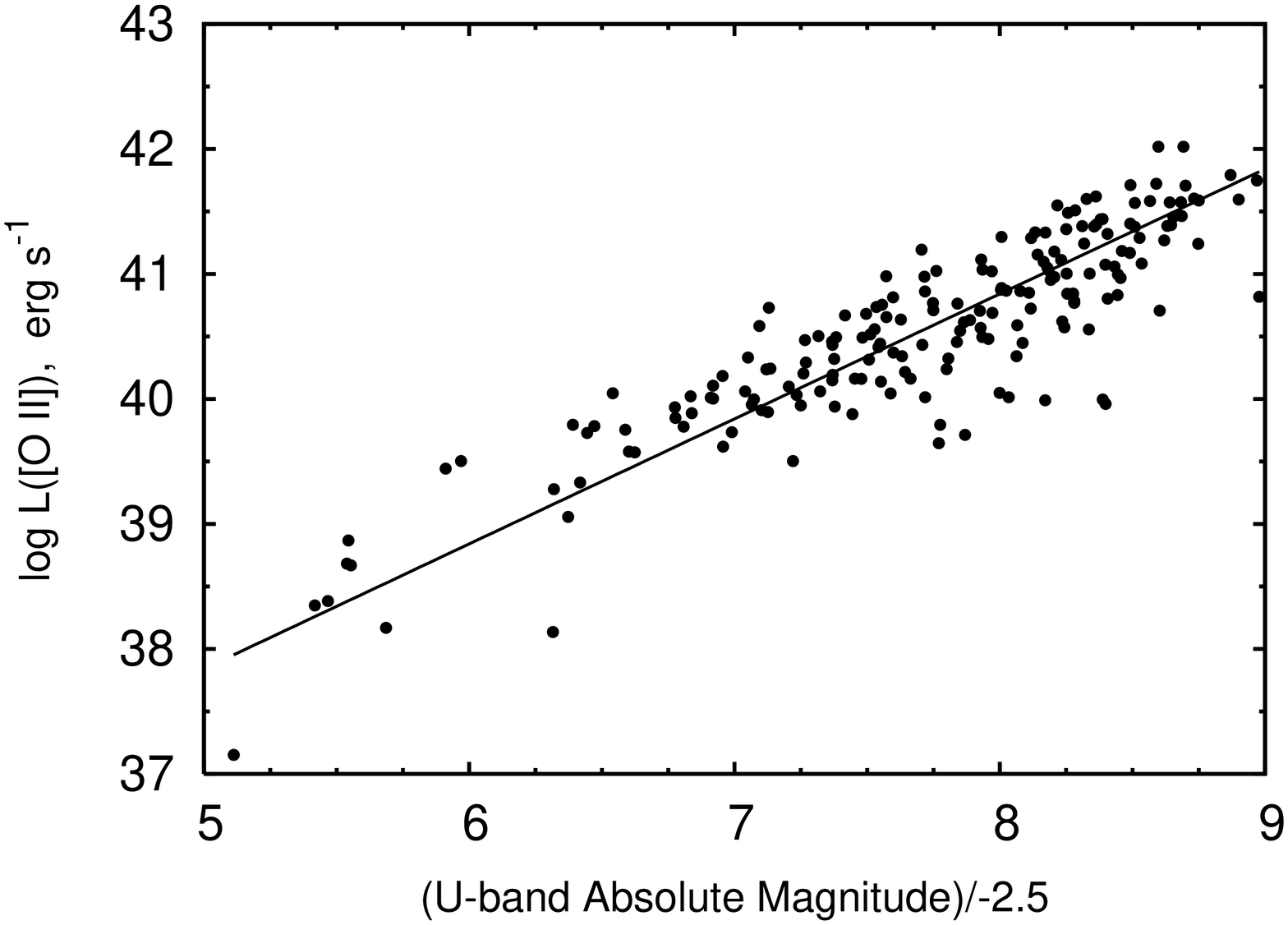,angle=-90.,width=6.5in}}
\noindent{\scriptsize
\addtolength{\baselineskip}{-3pt} 
\hspace*{0.3cm}Fig.~4.\ U-band flux versus \OII\ flux for the Moustakas \& Kennicutt (2006)
sample of local starforming galaxies.  The \OII\ spectra are unique in that the spectrograph 
was scanned over a large fraction of the galaxy.  The solid line is the best-fit line of unity slope, 
with coefficients given in eqn.~2; this relation lets U-band flux be used as a proxy for \OII.
\vspace*{0.2cm}
\addtolength{\baselineskip}{3pt}
}
\end{figure*}

\begin{figure*}
\vspace*{0.1cm}
\hbox{~}
\centerline{\psfig{file=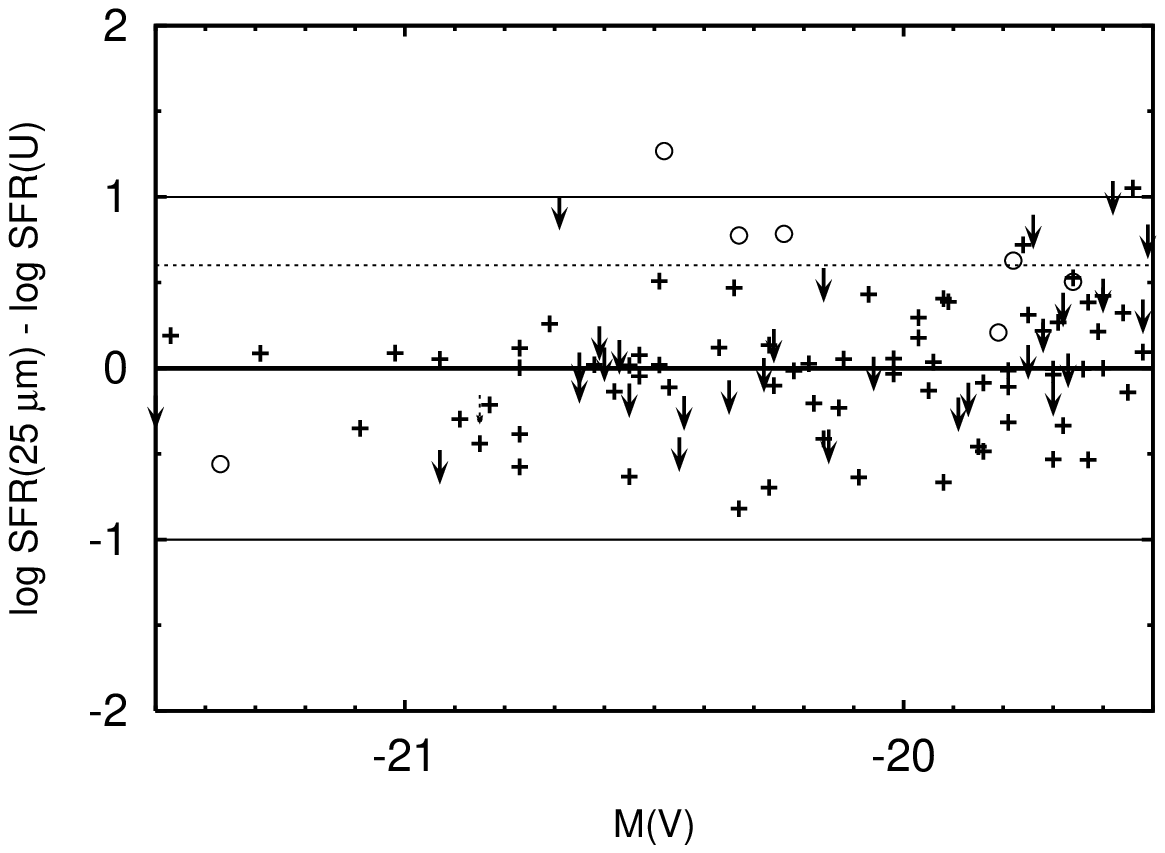,angle=0.,width=6.2in}}
\noindent{\scriptsize
\addtolength{\baselineskip}{-3pt}  
\hspace*{0.3cm}Fig.~5.\ Comparison of U-band and 25\micron--derived star formation rates
for the local volume--limited sample, using the \OII vs. U-band calibration (Fig. 4).  
The typical luminous ($M_V < -19.5$) starforming galaxy in the local sample has 
SFR(IR)/SFR(U) $\sim$ 1, in other words, the optical measurement of the star formation rate 
is consistent with the SFR determined from the mid-IR.  This suggests that dust obscuration is 
moderate for a typical present-epoch spiral or irregular galaxies.  Only a few galaxies
show significantly enhanced SFR(IR)/SFR(U) $\sim$3, indicative of  a high level 
of dust obscuration, whereas the {\it average} SFR(IR)/SFR(OII) $\sim$4 for the distant galaxy 
sample with the starburst signature.  Open symbols represent known AGNs, which are
excluded because their mid-IR fluxes are not mainly a product of star formation.  The errors in 
this plot are dominated by extrapolation to total U-band magnitudes, which are typically 20\% or
less.

\vspace*{0.2cm}
\addtolength{\baselineskip}{3pt}
}
\end{figure*}

U-band absolute magnitudes were converted to \OII\ luminosities using eqn.~2, and from that to 
SFR using the Kennicutt (1998) relation.  IRAS 25\micron\ luminosities were converted to L(8-1000\micron)
using the relation we fit to the Chary \& Elbaz (2001) models:

log L(8-1000\micron) (Lsun) = 0.89942*(log L$_{\nu}^{25}$ (Lsun) + 1.77782  (3)

\noindent  and from there to SFR using the relation from Kennicutt (1998).

\noindent To make the most straightforward comparison to the A851 sample, in Figure 5 we plot 
log SFR(IR) -- log SFR(U) versus $M_V$ for the local sample.  For local galaxies corresponding to 
the absolute magnitude distribution of our A851 sample,  $-22 < M_V < -19.5$, we find that, on average, 
the $SFR(IR) \ls SFR(U)$.  That is, we find few galaxies comparable to the SFR(IR)/SFR(\OII) $\sim$ 4 
that we found to be typical of the e(a) galaxies in A851, those identified as starbursts.  Rather, we find 
the local sample to cluster around SFR(IR)/SFR(U) $\sim$ 1, values that we found for the e(c) 
galaxies in A851 --- galaxies without starbursts.  There appears to be a rising trend in SFR(IR)/SFR(U) 
for the faintest objects in the local sample, but the IRAS sensitivity limits begin to be important, and these
are, at any rate, fainter than the galaxies in our A851 sample.

As we suggested earlier, finding SFR(U) $\approx$ SFR(\24m) for the large local sample and
the relatively small e(c) sample in A851 supports the claim that we are dealing with the same
kinds of objects, and it also indicates that our \24m SFR diagnostic is reasonably good.   

From our small but complete sample, we can see that for typical, present-epoch, starforming 
galaxies, star formation is only moderately obscured by dust. While it is not hard to find galaxies 
that are significantly dust-obscured today, they account for only a few percent of the local population 
(see, for example, Rieke \& Lebofsky 1978). For the Morphs ten-cluster sample, Dressler \etal (1999) 
found a typical fraction of e(a) galaxies  of $10-20\%$.  Assuming the association of dusty starbursts 
with e(a) spectra we find in A851 is representative of this larger sample, this points to a factor of $\sim$4 
decrease in the number of dusty starbursts in the last 4 Gyr, a relatively short time in cosmic terms.

\section{The star formation histories in A851 galaxies}

The next step should be to analyze the spectra of our galaxies to see what the heuristic 
distinctions `starburst, post-starburst, and `continuously starforming' really mean in terms of 
star formation histories.  Our data for this study is minimal for this purpose, but here we 
develop the tools in anticipation of the much larger, higher quality data set to come.  

Our A851 galaxies have on-average low S/N spectra, so we follow the technique of 
Dressler \etal\ (2004) to increase the S/N substantially and create representative templates 
of each class.  We use the spectrophotometric model of Fritz \etal\ (2007) to analyze the 
composite spectra in order to derive approximate star formation histories of the various 
classes.  The model performs a simultaneous fit to the stellar continuum and to a number of 
emission and absorption features in each observed composite spectrum, as well as to
the total IR luminosity, as derived by the \24m flux (Section 4.3).   For the A851 composites
we substituted the broad-band fluxes in place of the optical continuum shape, due to uncertainties 
in the relative flux calibration of these spectra.  

The program measures equivalent widths of absorption and emission features in
the modeled, synthetic spectra in the same way as is done for the observed spectrum. 
This requires that, in performing the fit, spectra representing a single stellar population,
or SSP, are degraded to the resolution of our observed spectra, and also that spectral 
regions with sky-subtraction problems 
are excluded.  The result is a sum of single stellar populations (SSPs) that
minimize a $\chi^2$ function that measures the differences between the model 
and the observed values of the observables.  The fitting algorithm is discussed in 
detail in Fritz \etal. 

Dust obscuration is allowed to vary with stellar age, to account for the fact that younger 
stars are generally more obscured than older ones.  Three metallicities --- solar, Z=0.05, 
and Z=0.004  --- are explored when searching for the best fit.  For each value of SSP 
metallicity, we perform 11 different fits of the same observed spectrum changing both
the initial point in the parameter space and the seeds of the random number generator, 
to evaluate the robustness of our results and the ``spread" in parameter space (i.e., in 
star formation history) of all acceptable solutions.  We choose the metallicity yielding the 
best $\chi^2$. ÊAmong the 11 models of that metallicity, we choose as reference model 
the one that has the median value of the total stellar mass: all the quantities quoted in this 
paper, such as masses, star formation rates, and extinction values, refer to this model. ÊError 
bars are computed as the half-difference between the two models (among the 11) with the 
most disparate values of total mass from the reference model. Details on the error 
determination for the SFRs computed by our model can be found in Fritz et al. (2007). The model 
employed here uses the new MILES observed stellar library from Sanchez-Blazquez \etal 
(2006) and a Salpeter IMF with stellar masses in the range $0.15 \le M/\Msun \le 120$.

Finally, since not all galaxies are detected by Spitzer-MIPS, we explored two cases: one in 
which the non-detections correspond to an IR luminosity = 0, and one adopting as IR 
emission the upper limit of our Spitzer data, $SFR = 3 ~\Msun yr^{-1}$. These two cases 
should bracket the possible range of star formation rates acceptable, and allow us to 
assess the uncertainty in the conclusions due to the unmeasured IR fluxes.

By determining the contribution of SSPs of different ages, the model yields a star 
formation history. That is, the mass formed at each cosmic time up to the epoch of 
observation, the extinction by dust and the total stellar mass for each composite spectrum.  
The synthetic spectra computed by the model are compared to the input composite-spectra
observations in Figure 6.  In the next sections we discuss each of these classes and
the SFR histories that have been derived.

\subsection{Post-starburst spectra}

Post-starburst spectra with and without a Spitzer-MIPS detection have been modeled separately  
in order to compare their star formation histories.  In the following discussion we will use
the term ``observation epoch" --- OE --- to refer to the properties of galaxies as observed
at the lookback time, for this case, approximately 4 Gyr ago.

The Spitzer-detected, post-starburst (k+a) composite spectrum is best fitted by a population
of age $\tau > 6 \times 10^9$ years that comprises 70\% of the OE galaxy stellar mass, 
followed by a strong burst occurring over $2-5 \times 10^8$ yr before OE that produced most of the 
remaining 30\%.  The IR detection by itself indicates residual continuing star formation 
at the OE (Figure 7).  The average galaxy stellar mass for this type was $M = 1.50 \times 10^{11} \, \Msun$.
The average SFR per galaxy was about 16 $\Msun~yr^{-1}$ over the galaxy's history, but during the last 
$\tau \ls 6 \times 10^8$~ yr, the SFR was considerably higher, 72  $\Msun~yr^{-1}$. More recently, over the 
last $2 \times 10^7$~yr before OE, the rate was $9-10~ \Msun~yr^{-1}$, significant in absolute terms, but 
substantially less to the extended high rate of star formation that preceded it.  To summarize in words, the 
IR-detected post-starburst galaxies are galaxies that have experienced a substantial burst of star 
formation --- significantly above the long-term past average --- followed by a level of ongoing star formation 
that is significant in absolute terms, but a sharp decline from the previous burst.\footnote{Though the 
equivalent widths of the spectral features and the broad-band continuum fluxes can be reproduced, 
our model fails to account for the broadness of the Balmer lines in Spitzer-detected a+k/k+a, 
as can be seen in Figure 6 (second spectrum from the top).  We found no single SSP or combination 
that was able to  account for this. We simulated the effects of an high velocity dispersion of 250 
$\rm km \, s^{-1}$ {\it for the galaxy},  by convolving our SSPs with a Gaussian.  The top spectrum 
of Figure 6 shows the best fit star formation history and rate unchanged with respect to the unbroadened 
fit, however, the line width is now reproduced. The broadening of the lines due to high galaxy velocity 
dispersions is consistent with the high average galaxy stellar mass found by both types of fits, 
$M = 1.5 \times 10^{11} \, \Msun$.  This mass agrees well with the value expected for early-type 
galaxies:  assuming $M/L = 4$ (at the low end of the range for early type galaxies, and thus appropriate 
for this post-starburst example), the predicted velocity dispersion from a modern determination of the 
Faber-Jackson relation in the Coma cluster (Matkovic and Guzman 2007) is $\sigma = 230~\kms$.} 

Our model for the {\it IR-undetected} post-starburst (k+a) spectrum confirms a substantial burst that 
is followed by a low-to-zero level of residual star formation (depending on how the IR non-detections
are handled).  The average SFR per galaxy in this group is $\sim12~\Msun~yr^{-1}$ over the 
typical galaxy's history, rising by a factor of 3.3 (1.4) to 39 (17) $\Msun~yr^{-1}$ (the first number 
corresponds to setting IR flux $=0$, while the number in parentheses corresponds to setting the
IR flux $=$ upper limit) during the last $6 \times 10^8$\,yr, and  0.0 (0.9)  solar masses per year during 
the last $2 \times 10^7$\,yr.  In words, the burst of star formation represents a significant rise over the past 
average, and there is no significant star formation at the epoch of observation --- a definite post-starburst.

The average galaxy stellar mass for this class is somewhat lower than for the Spitzer-detected 
post-starbusts, yielding 1.1 $\times 10^{11} \, \Msun$ regardless of the IR limits adopted, while
the mass fraction formed during the last $6 \times 10^8$~yr varies significantly depending on whether 
no IR emission or IR upper limits are included in the fit, $\sim 20\%$ versus 9\%.

The post-starburst galaxies with and without a Spitzer detection have in common a history 
where a significant fraction, $10-30$\% of the galaxy mass has formed in a recent burst with 
exceptionally high star formation rates. A lower limit for the SFR {\it during} the burst can be
estimated from the average SFR during the last $6 \times 10^8$\,yr, thus SFR $> 20-70~ 
\Msun \, yr^{-1}$.


\begin{figure*}
\vspace*{0.1cm}
\hbox{~}
\centerline{\psfig{file=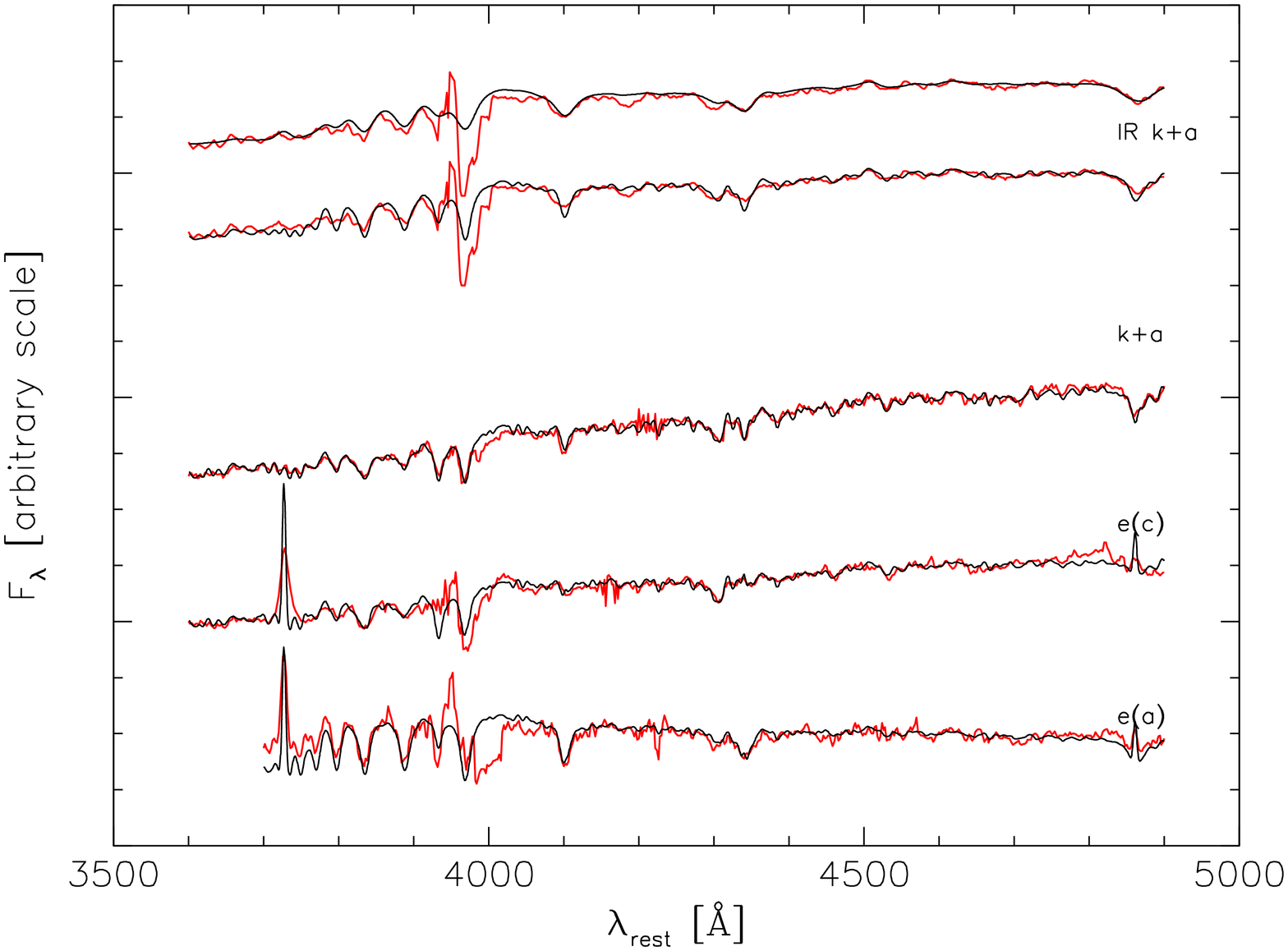,angle=-90.,width=7.0in}}
\noindent{\scriptsize
\addtolength{\baselineskip}{-3pt} 
\hspace*{0.3cm}Fig.~6.\  Observed composite spectra (red, thick line) compared to models  (overlaid.
thin black line). From top to bottom: post-starburst galaxies with and without IR emission, of starforming galaxies 
and dusty starbursts. Broad-band galaxy photometry, rather than the spectroscopic continuum, was used in the 
fitting procedure.  The fits were to these broad band colors and the equivalent widths of the emission and 
absorption lines, not the shape of the lines.  The line shape fits are acceptable except for the infrared 
post-starburst spectrum (IR k+a), which has significantly broader Balmer absorption lines than the model. 
The bottom of the two templates for IR-detected k+a galaxies shows a poor fit to the width of the Balmer
absorption lines.  The top spectrum shows that we are able to reproduce the Balmer line profiles by smoothing 
the model fit to correspond to a velocity dispersion of $250$ \kms, which is appropriate for elliptical galaxies of 
the mass of the average k+a in the composite.

\vspace*{0.2cm}
\addtolength{\baselineskip}{3pt}
}
\end{figure*}


\begin{figure*}
\vspace*{0.1cm}
\hbox{~}
\centerline{\psfig{file=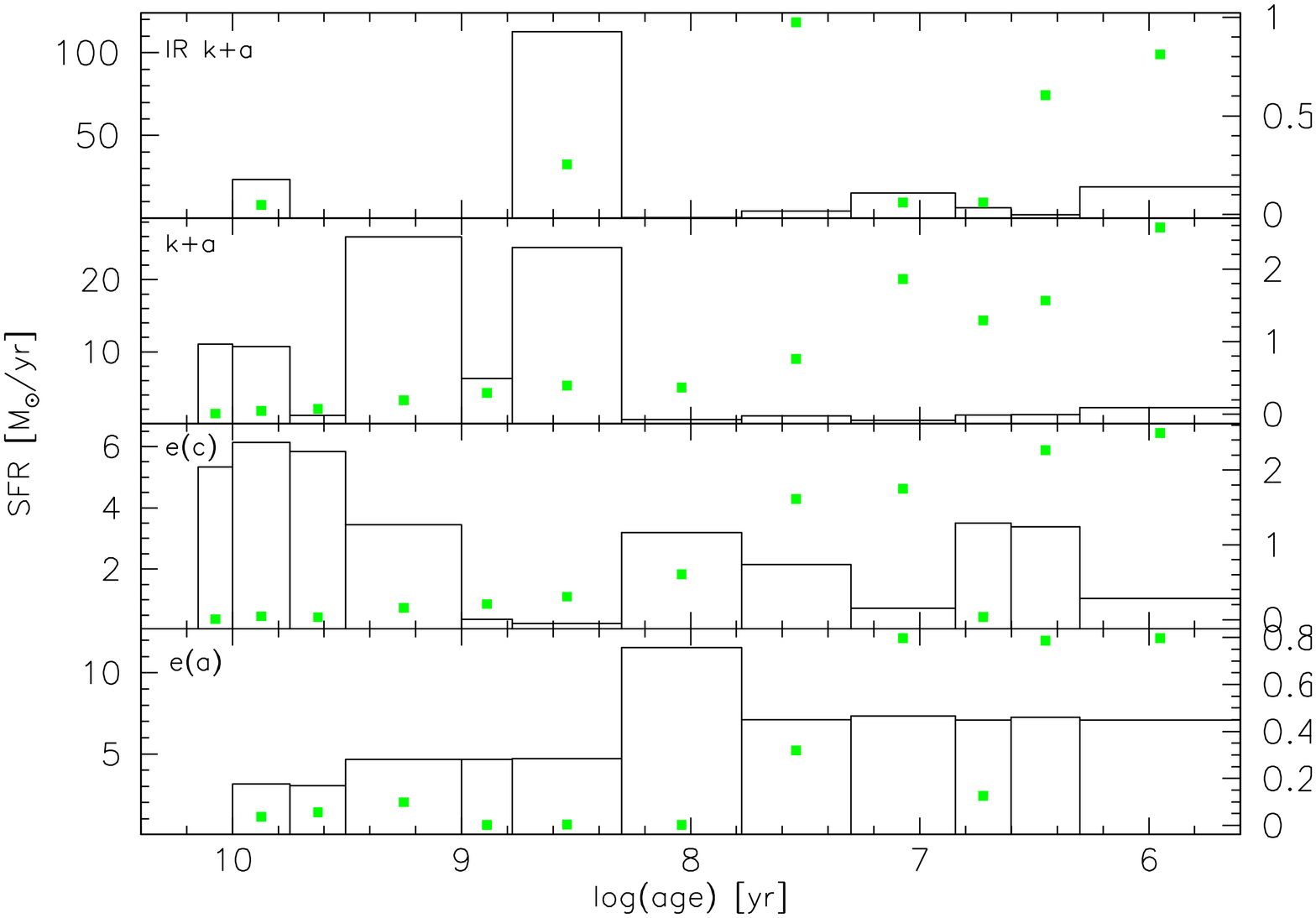,angle=-90.,width=7.0in}}
\noindent{\scriptsize
\addtolength{\baselineskip}{-3pt}  
\hspace*{0.3cm}Fig.~7.\  Star formation rate as a function of SSP age for the best-fit models
of post-starburst galaxies with IR emission (IR k+a), post-starburst galaxies without IR emission 
(k+a), starforming galaxies [e(c)], and dusty starbursts [e(a)].  SFRs are in $\Msun yr^{-1}$, averaged
over the number of galaxies of each spectral class.  Age = 0 corresponds to the epoch of
observation.  Green dots indicate the value of E(B-V) for each stellar population, with the scale shown
on the right.

\vspace*{0.2cm}
\addtolength{\baselineskip}{3pt}
}
\end{figure*}

\subsection{Emission-line spectra}

The star formation history that best fits the e(c) ``continuously starforming galaxy" composite 
spectrum is markedly different from the two similar starburst histories described above.  The 
average SFR per galaxy in this group is about 7.5 $\Msun \, yr^{-1}$ over the typical galaxy's history, 
between 1.6 (0.6) $\Msun \, yr^{-1}$ (null IR flux or upper limit) during the 
last $6 \times 10^8$\,yr and 2.1 (1.5)  $\Msun \, yr^{-1}$ during the last $2 \times 10^7$\,yr 
before OE.  Recent and ongoing SF levels are low, significantly lower than the all-time past 
average.  It is important to stress that the variations of SFR over a single stellar population should
not be considered significant, as the model cannot have such detailed time resolution.
The overall SFR trends, as quantified in the broad age intervals of which will be presented inTable 2, are 
significant, but smaller intervals with higher or lower SFRs are not.  For example, in the case of the e(c)
galaxies, we note that the observed composite spectrum is noisier than in the post- or dusty- starburst 
composites, due both to the small number of objects (7) and their relative faintness.    The full star formation
history shown in Figure 7 for the e(c) shows many ups and downs, but these are not statistically
significant.   The overall star formation history --- with these data --- can best be interpreted as strong in the 
first few Gyr ($z \sim 2$) and steadily declining thereafter.  Unlike the post-starbursts discussed above, and 
the dusty-starbursts discussed next, there are no important bursts for many Gyr before OE.

The stellar mass determined by the model for the average galaxy in the e(c) group is around 
$7 \times 10^{10} \, \Msun$, lower than the either of the post-starburst types discussed above. 

The average galaxy mass is even lower for the fit of the e(a) ``dusty-starburst" spectrum: for
the typical galaxy in this class $M = 2.5 \times 10^{10} \, \Msun$.  The model history of the dusty 
starbursts is consistent with an all-time average SFR per galaxy of 2.7 (2.8) (null IR or upper limit), 
while the recent and ongoing SFRs are 6.4 (6.6) during the last $6 \times 10^8$\,yr and 6.9 (7.3) during 
the last $2 \times 10^7$\,yr. The recent star formation in these galaxies is therefore more than twice 
the all-time average.  The rise in SFR compared to the $\sim10^9$ years immediately preceding the
burst might actually be more than a factor-of-two, because most massive galaxies show histories of
monotonically declining SFR over time, in other words, using the ``all-time" average is
likely to be an upper limit for the period immediately preceding the burst.  The modeling cannot, without
very high S/N spectra, resolve this difference, but a more accurate assessment of the rise in SFR compared
to the time immediately preceding the burst, would be important in understanding the actual physical
conditions and mechanisms that might be responsible.

In summary, the e(a) galaxies have had starbursts like the k+a types, but these starbursts -- substantially 
hidden b dust --- are continuing at the OE.  

Table 2 contains a summary of the model results for each of the different spectral classes, including 
the SFR per galaxy averaged over the last $2\times 10^7$, $6\times 10^8$ and over the age 
of the universe at $z=0.4$ (``all-time"), $\tau = 9.2 \times 10^9$\,yr). Also listed is the number 
of galaxy spectra which contributed to each composite spectrum. The classification into
spectral classes has clearly identified galaxies with markedly different stellar histories. 
The k+a galaxies are once more confirmed to be post-starburst galaxies, with or without 
(depending on IR detection) residual star formation activity.  Among the emission-line galaxies, 
the star formation activity has declined with time in e(c) galaxies without a detectable burst, 
while in e(a) galaxies there is a mild-to-moderate burst of star formation at the OE.

The models are constrained to reproduce the FIR luminosity inferred from the mid-IR luminosity, not the SFR
as indicated from the \24m measurements (section 4.3).   Nevertheless, we find good agreement between
what the modeling returns for the SFR over the time bin $2 \times 10^7$ yr and our ``direct" determination of
the SFR from the mid-IR.  The agreement is not expected to be exact because (1) the FIR (from mid-IR) flux
is one of many constraints in the models, and (2) the two paths incorporate different assumptions about the 
initial mass function and the timescales for the relevant calculations.  

We summarize this section by noting that the star formation histories derived with the spectral
modeling of star formation are consistent with the simple picture presented in Section 5 where we
compared \24m and [O II] fluxes to distinguish the histories of galaxies of the different spectral 
classes.  The results of this small, first study using Spitzer-MIPS data are consistent with the idea that e(a) 
and e(b) galaxies are active starbursts, k+a galaxies are the post-starburst phase of those galaxies,
and e(c) galaxies are those that have not experienced any starbursts in the past few billion years.
The numbers of cases and spectra are sufficiently small that we must consider this a preliminary 
step toward confirming the Poggianti \etal (1999) interpretation of galaxy spectra in 
intermediate-redshift clusters.  For example, the e(a) galaxies studied here have a mass 2-3
times smaller than the two types of post-starbursts.  This could easily be the result of the small
sample sizes, but if it were to be confirmed with the much larger study we have begun, this would
provide a puzzling distinction that would not fit with the simple model.  

Cycle GO-4 observations of  27.5 hrs (PID 40387, PI A. Dressler) have been awarded to 
extend the coverage of Abell 851, and to observe two additional clusters at similar redshift,
for which large spectroscopic samples of field and cluster galaxies have been obtained.  
These data should allow further investigation of obscured star formation
in this cluster and its relevance to other intermediate-redshift  galaxy populations.

\section{Conclusions}

Our study of optical spectroscopy and Spitzer-MIPS observations of a spectroscopic sample of cluster 
members of A851 at $z = 0.42$ has yielded the following results: (1) Detection with MIPS at \24m
is a strong function of spectral type: almost all passive galaxies are undetected, and most normal
starforming galaxies (type e(c), weak \Hd absorption) are also undetected. However, some 
post-starburst galaxies (strong \Hd) and {\it most} e(a) galaxies (strong \Hd {\it and} \OII emission) 
are detected.  The non-detection of e(c) galaxies as compared to the detection of most e(a) galaxies
occurs across the full range of luminosity sampled --- in fact, the e(c)'s are systematically more 
massive in this sample --- so this difference in detection is not a selection effect. (2) For the
e(a) galaxies, the SFR derived from \24m is typically several times that inferred from \OII, indicating a
substantial amount of dust obscuration of ongoing star formation in these systems.  Ongoing
star formation is seen in some galaxies classified as post-starburst, for which 
SFR(\24m)/SFR(\OII) is likely to be as much as a factor of 10, but this it represents a significant 
decline from the burst that came before.  (3) The high rates of star formation
in e(a) galaxies --- factors of 2-3 above the past average for these systems --- qualifies them as 
moderate starbursts capable of accounting for the post-starbursts seen in A851, although for
the specific, small samples we have here, there is a factor of 2--3 lower mass associated with the 
e(a) compared to the k+a galaxies. (4) A small sample of present-day starforming 
galaxies shows that these typically have SFR(IR)/SFR(optical) $\sim$ 1.  We suggest that these 
galaxies are like the e(c) galaxies in A851, which would be consistent  with these objects 
having \24m flux below the detection limit of these observations.

The assembly of these observations supports the model in which the e(a) galaxies and even 
some post-starburst galaxies are actually dust-obscured starbursts, with a factor of $\sim$4
of the star formation hidden.  As we will show in a forthcoming paper, the significant fraction
of e(a) spectra among field galaxies at intermediate-redshift suggests that starbursts are also
relatively common among field galaxies at intermediate-redshift.  The rarity of such objects in 
the present-epoch universe indicates that star formation in ordinary spirals was more 
bursty --- more variable ---  as recently as 4 Gyr ago.  The extent and magnitude of this phenomenon 
will become increasingly clear as further Spizter-MIPS observations become available for 
intermediate-redshift clusters and the field.  

\section{Acknowledgments}

The authors thank the referee for a careful review of the manuscript and many helpful suggestions.
Dressler and Oemler acknowledge the support of the NSF grant AST-0407343.  All the authors
thank NASA for its support through NASA-JPL 1310394.  Jane Rigby is supported by a Spitzer Space
Telescope Postdoctoral Fellowship.  Partial support was also provided through contract 1255094
from JPL/Caltech to the University of Arizona.

%
%
\begin{deluxetable}{lrr}
\tablecaption{Detection Rate of Spectral Types}
\tablehead{\colhead{Type} & \colhead{Total} & \colhead{Detected}}
\startdata
k		&	41	&	2\\
e(c)	&	7	&	2\\
e(a)\tablenotemark{a}	&	12	&	9\\
e(b)	&	3	&	1\\
k+a\tablenotemark{a}	&	13	&	4\\
a+k	&	5	&	2\\
e(n)	&	2	&	2\\
\enddata
\tablenotetext{a}{As explained in the text, two objects meeting the k+a criteria --- $\EWOII > -5$\AA\ and $\EWHd > 3$
--- but  with detected \OII are included in the e(a) rather than k+a numbers.}
\end{deluxetable}

%
%
\begin{table}
\begin{center}
{
\caption{Model summary.\label{tbl1}}
\vspace{1.0cm}
\begin{tabular}{lcccc}
\tableline\tableline
&&&& \\
                  & IR-det. k+a & Undet. k+a & e(c) & e(a) \\
\tableline
&&& \\
$N_{obj}$			&	 6	 &   12   &   8  &   19\\
``all-time"          &  $16 \pm 3$  &  $12 \pm 2$    &  $ 7 (7) \pm 3$  &  $3 (3) \pm 0.8$ \\
$6\times 10^8$yr    &  $72 \pm 7$   &  $39 (17) \pm 5 (1)$ &  $1 (2) \pm 2$ &  $6 (7) \pm 2$ \\
$2\times 10^7$yr    &   $9 \pm 1$ &  $1.1 \pm 0.1$    &  $1 (2) \pm 1$  &  $7 (7) \pm 3$ \\
&&&&\\
SFR(\24m)     &   12.1  &  $<$3.0  & 1.5  & 7.7 \\
&&&&\\
Mass &  $1.5 \pm 0.3  \times 10^{11}$ & $1.1 \pm 0.1 \times 10^{11}$ &  $7 \pm 0.6 \times 10^{10}$
& $2.5 \pm 0.6 \times 10^{10}$ \\
\tableline
\end{tabular}
}
\tablecomments{Average SFR ($M_{\odot} \, yr^{-1}$) per galaxy of each spectral 
class, as determined in the models. The SFR is averaged over the last $2\times 10^7$,
$6\times 10^8$ and over the age of the Universe at $z=0.4$ --- ``all-time", $9.2\times 
10^9$ yr. The values in parentheses come from adopting the \24m upper limit as
opposed to the null value, in case of non-detection.  The SFR(\24m) values, actual 
measurements, are to be compared to the $2 \times 10^7$ yr SFR from the model, as
explained in the text. The last row contains the average galaxy stellar mass in solar 
masses.}
\end{center}
\end{table}

\vfill\eject
\clearpage


\end{document}